\documentclass[twoside,fleqn]{article}
\usepackage{espcrc2}
\newcommand{\p}{\partial}

\newcommand{\AmS}{{\protect\the\textfont2
  A\kern-.1667em\lower.5ex\hbox{M}\kern-.125emS}}

\title{Dimension two gluon condensates in a variety of gauges and a gauge invariant Yang-Mills action with a mass}
\author{D.~Dudal
\thanks{david.dudal@ugent.be. D.~Dudal is a postdoctoral fellow
of the \emph{Special Research Fund} of Ghent University.}
\address[Gent]{Ghent
University, Department of Mathematical Physics and Astronomy,
Krijgslaan 281-S9, B-9000 Gent, Belgium}, M.A.L.
Capri\address[Rio]{UERJ - Universidade do Estado do Rio de Janeiro,
Rua S\~{a}o Francisco Xavier 524, 20550-013 Maracan\~{a}, Rio de
Janeiro, Brasil}, J.A. Gracey\address{Theoretical Physics Division,
Department of Mathematical Sciences, University of Liverpool, P.O.
Box 147, Liverpool, L69 3BX, United Kingdom}, V.E.R.
Lemes\addressmark[Rio] R.F. Sobreiro\addressmark[Rio], S.P.
Sorella\thanks{Work supported by FAPERJ, Funda{\c c}{\~a}o de Amparo
{\`a} Pesquisa do Estado do Rio de Janeiro, under the program {\it
Cientista do Nosso Estado}, E-26/151.947/2004. }\addressmark[Rio],
H.~Verschelde\addressmark[Gent] }
\begin{document}

\begin{abstract}
We give a short overview of our work concerning the dimension two
operator $A^2$ in the Landau gauge and its generalizations to other
gauges. We conclude by discussing recent work that leads to a
renormalizable gauge invariant action containing a mass parameter,
based on the operator $F\frac{1}{D^2}F$.
\end{abstract}

\maketitle
\section{Introduction}
Recent years have witnessed a great deal of interest in the possible
existence of mass dimension two condensates in gauge theories, see
for example
\cite{Gubarev:2000eu,Gubarev:2000nz,Verschelde:2001ia,Kondo:2001nq,Boucaud:2001st,Dudal:2002pq,Dudal:2003by,Andreev:2006vy,Csaki:2006ji}
and references therein for approaches based on phenomenology,
operator product expansion, lattice simulations, an effective
potential and the string perspective. There is special interest in
the operator
\begin{equation}\label{m1}
    A^2_{\min}=(VT)^{-1}\min_{U\in SU(N)}\int d^4x
    \left(A^U\right)^2\,,
\end{equation}
since it is gauge invariant due to the minimization along the gauge
orbit. It should be mentioned that obtaining the global minimum is
delicate due to the problem of gauge (Gribov) ambiguities
\cite{Gribov:1977wm,Semenov}. As is well known, \emph{local} gauge
invariant dimension two operators do not exist in Yang-Mills gauge
theories. The nonlocality of (\ref{m1}) is best seen when it is
expressed as \cite{Lavelle:1995ty}
\begin{eqnarray}\label{m2}
&&A_{\min }^{2} =\frac{1}{VT}\int d^{4}x\left[ A_{\mu }^{a}\left( \delta _{\mu \nu }-\frac{%
\partial _{\mu }\partial _{\nu }}{\partial ^{2}}\right) A_{\nu
}^{a}\right.\nonumber\\&-&\left.gf^{abc}\left( \frac{\partial _{\nu
}}{\partial ^{2}}\partial A^{a}\right) \left( \frac{1}{\partial
^{2}}\partial {A}^{b}\right) A_{\nu }^{c}\right] +\ldots\,.
\end{eqnarray}
The relevance of the condensate $\left\langle
A^2\right\rangle_{\min}$ was discussed in
\cite{Gubarev:2000eu,Gubarev:2000nz}, where it was shown that it can
serve as a measure for the monopole condensation in the case of
compact QED.
\section{Measurement of $\left\langle A^2\right\rangle_{\min}$}
All efforts so far concentrated on the Landau gauge $\p A=0$. The
preference for this particular gauge fixing is obvious since the
nonlocal expression (\ref{m2}) reduces to a local operator, more
precisely
\begin{equation}\label{m3}
    \p A=0\Rightarrow A^2_{\min}=A^2\,.
\end{equation}
In the case of a local operator, the Operator Product Expansion
(OPE) becomes applicable, and consequently a measurement of the soft
(infrared) part $\left\langle A^2\right\rangle_{OPE}$ becomes
possible. Such an approach was followed in e.g.
\cite{Boucaud:2001st} by analyzing the appearance of $\frac{1}{q^2}$
power corrections in (gauge variant) quantities like the gluon
propagator or strong coupling constant, defined in a particular way,
from lattice simulations. Let us mention that already two decades
ago attention was paid to the condensate $\left\langle
A^2\right\rangle$ in the OPE context \cite{Lavelle:1988eg}.

The condensate $\left\langle A^2\right\rangle_{OPE}$ can be related
to an effective gluon mass, see e.g. \cite{Kondo:2001nq}. Effective
gluon masses have found application in some phenomonological
studies. Also lattice simulations of the gluon propagator revealed
the need for massive parameters
\cite{Langfeld:2001cz,Amemiya:1998jz,Bornyakov:2003ee}.

A more direct approach for a determination of $\left\langle
A^2\right\rangle$ in the Landau gauge was presented in
\cite{Verschelde:2001ia}. A meaningful effective potential for the
condensation of the \emph{Local Composite Operator} (LCO) $A^2$ was
constructed by means of the LCO method. This is a nontrivial task
due to the compositeness of the considered operator. We consider
pure Euclidean SU(N) Yang-Mills theories with action
\begin{eqnarray}\label{m3}
 S_{YM}&=&\int d^4 x \frac{1}{4}F_{\mu\nu}^a F_{\mu\nu}^a
+S_{gf}\,,\nonumber\\
S_{gf}&=& \int d^4x\left(b^a\p_\mu A_{\mu}^a+\overline{c}^a\p_\mu
D_\mu^{ab}c^b\right)\,.
\end{eqnarray}
We couple the operator $A^2$ to the Yang-Mills action by means of a
source $J$,
\begin{eqnarray}\label{m4}
 S_{J}&=&S_{YM}+\int d^4x\left(\frac{1}{2}JA_\mu^aA_\mu^a-\frac{1}{2}\zeta
 J^2\right)\,.
\end{eqnarray}
The last term, quadratic in the source $J$, is necessary to kill the
divergences in vacuum correlators like $\left\langle
A^2(x)A^2(y)\right\rangle$ for $x\to y$, or equivalently in the
generating functional $W(J)$, defined as
\begin{equation}\label{m5}
    e^{-W(J)}=\int[\mbox{fields}] e^{-\int d^4x S_J}\,.
\end{equation}
The presence of the LCO parameter $\zeta$ ensures a homogenous
renormalization group equation for $W(J)$. Its arbitrariness can be
overcome by making it a function $\zeta(g^2)$ of the strong coupling
constant $g^2$, allowing one to fix $\zeta(g^2)$ order by order in
perturbation theory in accordance with the renormalization group
equation.

In order to recover an energy interpretation, the term $\propto J^2$
can be removed by employing a Hubbard-Stratonovich transformation
\begin{equation}\label{m6}
    1=\int\sigma
    e^{-\frac{1}{2\zeta}\left(\frac{\sigma}{g}+\frac{1}{2}A^2-\zeta J\right)^2}\,,
\end{equation}
leading to the action
\begin{eqnarray}\label{m7}
S&=&S_{YM}+S_\sigma\,,\nonumber\\
S_\sigma&=&\int
d^4x\left(\frac{\sigma^2}{2g^2\zeta}+\frac{1}{2g^2\zeta}g\sigma
A^2+\frac{1}{8\zeta}(A^2)^2\right)\,.\nonumber\\
\end{eqnarray}
A key ingredient for the LCO method is the renormalizability of the
operator $A^2$. It was proven in \cite{Dudal:2002pq} that $A^2$ is
renormalizable to all orders of perturbation theory, making use of
the Ward identities.

Starting from (\ref{m7}) it is possible to calculate the effective
potential $V(\sigma)$. The correspondence $\left\langle
\sigma\right\rangle=-g\left\langle A^2\right\rangle$ consequently
provides evidence for a nonvanishing dimension two gluon condensate
using an effective potential approach, if $\left\langle
\sigma\right\rangle\neq0$. It is clear from (\ref{m7}) that
$\left\langle \sigma\right\rangle\neq0$ induces an effective gluon
mass. $V(\sigma)$ was calculated to two loop order in
\cite{Verschelde:2001ia,Browne:2003uv}, and a nonvanishing
condensate is favoured as it lowers the vacuum energy. The ensuing
effective gluon mass was a few hundred MeV.

Before ending this section, we want to stress that the value
$\left\langle A^2\right\rangle_{LCO}$ has no clear connection with
$\left\langle A^2\right\rangle_{OPE}$. The former one is derived
from an effective potential calculated in perturbation theory, thus
a priori only reliable in the UV regime, while the latter one finds
it origin in the IR sector. Furthermore, the notion of a dynamical
gluon mass does not imply the existence of (physical) massive
gluons. Our results should rather be taken as giving evidence for
the appearance of nonperturbative mass parameters in the gluon
propagator, as also found by lattice simulations.

\section{$\left\langle A^2\right\rangle_{\min}$ beyond the Landau gauge?}
The question arises what can be said about the dimension two
condensate in a gauge other than the Landau gauge? As the operator
$A^2_{\min}$ is then clearly nonlocal, it falls beyond the
applicability of the OPE. It is also unclear how e.g.
renormalizability or an effective potential approach could be
established for nonlocal operators.

Nevertheless, in several other gauges, we have shown that other
dimension two, renormalizable, local operators exist. We generalized
the LCO method and showed that these operators condense and give
rise to a dynamical gluon mass, see Table 1 and
\cite{Dudal:2003by,Dudal:2003gu,Dudal:2003pe,Dudal:2004rx}.
\begin{table}
\begin{tabular}{|c|c|}
  \hline
  GAUGE&OPERATOR\\
  \hline\hline
  linear covariant &   $\frac{1}{2}A_\mu^a A_\mu^a$\\
  Curci-Ferrari & $\frac{1}{2}A_\mu^a
  A_\mu^a+\alpha\overline{c}^ac^a$\\
  maximal Abelian & $\frac{1}{2}A_\mu^{\beta} A_\mu^{\beta}+\alpha\overline{c}^{\beta}c^{\beta}$ \\
  \hline
\end{tabular}
\caption{Gauges and their renormalizable dimension two operator}
\end{table}
In the maximal Abelian gauge, it was found that only the
off-diagonal gluons $A_\mu^\beta$ acquire a dynamical mass, a fact
qualitatively consistent with the lattice results from
\cite{Amemiya:1998jz,Bornyakov:2003ee}.

We have been able to make some connection between the various gauges
and their dimension two operators by constructing renormalizable
interpolating gauges and operators \cite{Dudal:2004rx,Dudal:2005zr}.
\section{Search for a gauge invariant dimension two operator}
A disadvantage of the results so far is the explicit gauge
dependence of the used operator, and hence of the dynamically
generated mass. On one hand, we started looking for a gauge
invariant dimension two operator, which a fortiori needs to be
nonlocal. If we would like to have a consistent (calculational)
framework on the other hand, we should look for an operator that
could be localized by introducing a suitable set of extra fields.
From this perspective, $A^2_{\min}$ seems rather hopeless as it is a
infinite series of nonlocal terms, which would require an infinite
number of additional fields to localize. A much more appealing
operator is \cite{Capri:2005dy}
\begin{equation}\label{m10}
    F\frac{1}{D^2}F\equiv\frac{1}{VT}\int d^{4}xF_{\mu \nu }^{a}\left[
\left( D^{2}\right) ^{-1}\right] ^{ab}F_{\mu \nu }^{b}\,.
\end{equation}
Indeed, when we add this operator to the Yang-Mills action via
\begin{equation}
S_{YM}-\frac{m^{2}}{4}\int d^{4}xF_{\mu \nu }^{a}\left[ \left(
D^{2}\right) ^{-1}\right] ^{ab}F_{\mu \nu }^{b}\,,
\end{equation}
we can localize this action to
\begin{eqnarray}\label{m11}
&&S_{YM}+\int d^{4}x\left[\frac{im}{4}\left( B-\overline{B%
}\right) _{\mu \nu }^{a}F_{\mu \nu }^{a}\right.\nonumber\\
&+&\left.\frac{1}{4}\left( \overline{B}_{\mu \nu }^{a}D_{\sigma
}^{ab}D_{\sigma }^{bc}B_{\mu \nu }^{c}-\overline{G}_{\mu \nu
}^{a}D_{\sigma }^{ab}D_{\sigma }^{bc}G_{\mu \nu
}^{c}\right)\right]\,,\nonumber\\
\end{eqnarray}
at the cost of introducing a set of bosonic ($B_{\mu\nu}^a$,
$\overline{B}_{\mu\nu}^a$) and a set of fermionic ghost fields
($G_{\mu\nu}^a$, $\overline{G}_{\mu\nu}^a$), antisymmetric in their
Lorentz indices and belonging to the adjoint representation. The
local gauge invariance is respected with respect to
\begin{eqnarray}\label{m12}
\delta A_{\mu }^{a} &=&-D_{\mu }^{ab}\omega ^{b},,  \nonumber \\
\delta B_{\mu \nu }^{a} &=&gf^{abc}\omega ^{b}B_{\mu \nu }^{c} \,,
\delta \overline{B}_{\mu \nu }^{a} =gf^{abc}\omega
^{b}\overline{B}_{\mu \nu }^{c}\,,
\nonumber \\
\delta G_{\mu \nu }^{a} &=&gf^{abc}\omega ^{b}G_{\mu \nu }^{c} \,,
\delta \overline{G}_{\mu \nu }^{a} =gf^{abc}\omega
^{b}\overline{G}_{\mu \nu }^{c}\,.
\end{eqnarray}
Having found a reasonable classical action, we need to take a look
at the quantum properties of the action (\ref{m11}).

A first problem is the renormalizability. The action (\ref{m11}) as
it stands is not renormalizable \cite{Capri:2005dy}. Fortunately, we
were able to prove to all orders of perturbation theory the
renormalizability of the following slightly more general action
\begin{eqnarray}
  S_{phys} &=& S_{cl} +S_{gf}\;,\label{completeaction}\\
S_{cl}&=&\int d^4x\left[\frac{1}{4}F_{\mu \nu }^{a}F_{\mu \nu
  }^{a}+\frac{im}{4}(B-\overline{B})_{\mu\nu}^aF_{\mu\nu}^a
  \right. \nonumber\\&+&\left.\frac{1}{4}\left( \overline{B}_{\mu \nu
}^{a}D_{\sigma }^{ab}D_{\sigma }^{bc}B_{\mu \nu
}^{c}-\overline{G}_{\mu \nu }^{a}D_{\sigma }^{ab}D_{\sigma
}^{bc}G_{\mu \nu
}^{c}\right)\right.\nonumber\\
&-&\left.\frac{3}{8}%
m^{2}\lambda _{1}\left( \overline{B}_{\mu \nu }^{a}B_{\mu \nu
}^{a}-\overline{G}_{\mu \nu }^{a}G_{\mu \nu }^{a}\right)
\right.\nonumber\\&+&\left.m^{2}\frac{\lambda _{3}}{32}\left(
\overline{B}_{\mu \nu }^{a}-B_{\mu \nu }^{a}\right)
^{2}\right.\nonumber
 \\&+&\left.
\frac{\lambda^{abcd}}{16}\left(\overline{B}_{\mu\nu}^{a}B_{\mu\nu}^{b}-\overline{G}_{\mu\nu}^{a}G_{\mu\nu}^{b}%
\right)\right.\nonumber\\&&\left.\times\left( \overline{B}_{\rho\sigma}^{c}B_{\rho\sigma}^{d}-\overline{G}_{\rho\sigma}^{c}G_{\rho\sigma}^{d}%
\right) \right]\,,\\
S_{gf}&=&\int d^{4}x\;\left( \frac{\alpha }{2}b^{a}b^{a}+b^{a}%
\partial _{\mu }A_{\mu }^{a}+\overline{c}^{a}\partial _{\mu }D_{\mu
}^{ab}c^{b}\right)\,,\nonumber
\end{eqnarray}
in the class of linear covariant gauges. $\lambda^{abcd}$ is an
invariant rank 4 tensor coupling while $\lambda_{1}$ and $\lambda_3$
are mass couplings. The classical action $S_{cl}$ is still invariant
with respect to the gauge transformations (\ref{m12}). The gauge
fixed action itself enjoys a generalized BRST symmetry, generated by
the \emph{nilpotent} transformation
\begin{eqnarray}\label{m13}
s A_{\mu }^{a} &=&-D_{\mu }^{ab}c ^{b}\;,s c^{a} =\frac{g}{2}f^{abc}c^ac ^{b}\,,  \nonumber \\
 s B_{\mu \nu }^{a} &=&gf^{abc}c ^{b}B_{\mu \nu }^{c}
\;,s \overline{B}_{\mu \nu }^{a} =gf^{abc}c ^{b}\overline{B}_{\mu
\nu }^{c}\,,
\nonumber \\
s G_{\mu \nu }^{a} &=&gf^{abc}c ^{b}G_{\mu \nu }^{c} \,, s
\overline{G}_{\mu \nu }^{a} =gf^{abc}c ^{b}\overline{G}_{\mu
\nu }^{c}\,,\nonumber\\
s\overline{c}^{a} &=&b^a \,, s b^{a} =0\,,s^2=0\,.
\end{eqnarray}
In \cite{Capri:2005dy,Capri:2006ne}, we also presented various
renormalization group equations to two loop order, confirming the
renormalizability at the practical level. Various consistency checks
are at our disposal in order to establish the reliability of these
results, e.g. the gauge parameter independence of the anomalous
dimension of gauge invariant quantities or the equality of others,
in accordance with the output of the Ward identities in
\cite{Capri:2005dy}. Furthermore, we proved in \cite{Capri:2006ne}
the equivalence of the model (\ref{completeaction}) with the
ordinary Yang-Mills theory in the case that $m\equiv0$. An open
question is what the physical excitations are of the model in the
case that $m\neq0$. A useful tool in discussing this token will be
the nilpotent BRST charge associated to (\ref{m13}).
\section*{Acknowledgments.}
The Conselho Nacional de Desenvolvimento Cient\'{i}fico e
Tecnol\'{o}gico (CNPq-Brazil), the Faperj, Funda{\c{c}}{\~{a}}o de
Amparo {\`{a}} Pesquisa do Estado do Rio de Janeiro, the SR2-UERJ
and the Coordena{\c{c}}{\~{a}}o de Aperfei{\c{c}}oamento de Pessoal
de N{\'\i}vel Superior (CAPES) are gratefully acknowledged for
financial support.


\begin{thebibliography}{99}
\bibitem{Gubarev:2000eu}  F.~V.~Gubarev, L.~Stodolsky and V.~I.~Zakharov,
Phys.\ Rev.\ Lett.\ \textbf{86} (2001) 2220.

\bibitem{Gubarev:2000nz}  F.~V.~Gubarev and V.~I.~Zakharov,
Phys.\ Lett.\ B \textbf{501} (2001) 28.

\bibitem{Verschelde:2001ia}  H.~Verschelde, K.~Knecht, K.~Van Acoleyen and
M.~Vanderkelen, Phys.\ Lett.\ B \textbf{516} (2001) 307.

\bibitem{Kondo:2001nq}  K.~I.~Kondo,
Phys.\ Lett.\ B \textbf{514} (2001) 335.

\bibitem{Boucaud:2001st}  P.~Boucaud, A.~Le Yaouanc, J.~P.~Leroy,
J.~Micheli, O.~Pene and J.~Rodriguez-Quintero, Phys.\ Rev.\ D
\textbf{63} (2001) 114003.

\bibitem{Dudal:2002pq}  D.~Dudal, H.~Verschelde and S.~P.~Sorella,
Phys.\ Lett.\ B \textbf{555} (2003) 126.

\bibitem{Dudal:2003by}  D.~Dudal, H.~Verschelde, J.~A.~Gracey,
V.~E.~R.~Lemes, M.~S.~Sarandy, R.~F.~Sobreiro and S.~P.~Sorella,
JHEP \textbf{0401} (2004) 044.

\bibitem{Andreev:2006vy}
O.~Andreev, Phys.\ Rev.\ D {\bf 73} (2006) 107901.

\bibitem{Csaki:2006ji}
C.~Csaki and M.~Reece,  hep-ph/0608266.

\bibitem{Gribov:1977wm}
V.~N.~Gribov, Nucl.\ Phys.\ B {\bf 139} (1978) 1.

\bibitem{Semenov}  Semenov-Tyan-Shanskii and V.A. Franke, Zapiski Nauchnykh
Seminarov Leningradskogo Otdeleniya Matematicheskogo Instituta im.
V.A. Steklov AN SSSR, Vol. \textbf{120} (1982) 159. English
translation: New York: Plenum Press 1986.

\bibitem{Lavelle:1995ty}
M.~Lavelle and D.~McMullan, Phys.\ Rept.\  {\bf 279} (1997) 1.

\bibitem{Lavelle:1988eg}  M.~J.~Lavelle and M.~Schaden, Phys.\ Lett.\ B \textbf{208} (1988) 297.

\bibitem{Langfeld:2001cz}
K.~Langfeld, H.~Reinhardt and J.~Gattnar, Nucl.\ Phys.\ B {\bf 621}
(2002) 131.

\bibitem{Amemiya:1998jz}
K.~Amemiya and H.~Suganuma, Phys.\ Rev.\ D {\bf 60} (1999) 114509.

\bibitem{Bornyakov:2003ee}
V.~G.~Bornyakov, M.~N.~Chernodub, F.~V.~Gubarev, S.~M.~Morozov and
M.~I.~Polikarpov, Phys.\ Lett.\ B {\bf 559} (2003) 214.

\bibitem{Browne:2003uv}  R.~E.~Browne and J.~A.~Gracey,
JHEP \textbf{0311} (2003)  029.

\bibitem{Dudal:2003np}  D.~Dudal, H.~Verschelde, V.~E.~R.~Lemes,
M.~S.~Sarandy, R.~F.~Sobreiro, S.~P.~Sorella and J.~A.~Gracey,
Phys.\ Lett.\ B \textbf{574} (2003) 325.

\bibitem{Dudal:2003gu}  D.~Dudal, H.~Verschelde, V.~E.~R.~Lemes,
M.~S.~Sarandy, S.~P.~Sorella and M.~Picariello, Annals Phys.\
\textbf{308} (2003) 62.

\bibitem{Dudal:2003pe}  D.~Dudal, H.~Verschelde, V.~E.~R.~Lemes, M.~S.~Sarandy, R.~F.~Sobreiro,
S.~P.~Sorella, M.~Picariello and J.~A.~Gracey, Phys.\ Lett.\ B
\textbf{569} (2003) 57.

\bibitem{Dudal:2004rx}  D.~Dudal, J.~A.~Gracey, V.~E.~R.~Lemes,
M.~S.~Sarandy, R.~F.~Sobreiro, S.~P.~Sorella and H.~Verschelde,
Phys.\ Rev.\ D \textbf{70} (2004) 114038.

\bibitem{Dudal:2005zr}
D.~Dudal, J.~A.~Gracey, V.~E.~R.~Lemes, R.~F.~Sobreiro,
S.~P.~Sorella, R.~Thibes and H.~Verschelde, JHEP {\bf 0507} (2005)
059.

\bibitem{Capri:2005dy}
M.~A.~L.~Capri, D.~Dudal, J.~A.~Gracey, V.~E.~R.~Lemes,
R.~F.~Sobreiro, S.~P.~Sorella and H.~Verschelde, Phys.\ Rev.\ D {\bf
72} (2005) 105016.

\bibitem{Capri:2006ne}
M.~A.~L.~Capri, D.~Dudal, J.~A.~Gracey, V.~E.~R.~Lemes,
R.~F.~Sobreiro, S.~P.~Sorella and H.~Verschelde, Phys.\ Rev.\ D {\bf
74} (2006) 045008.

\end{thebibliography}
\end{document}